\documentclass{PoS}
\usepackage{amsmath,multicol,multirow,arydshln,graphicx,wrapfig}
\title{Effective Lagrangian Approach to Top Decay via Flavor Changing Neutral Current}

\ShortTitle{Effective Lagrangian Approach to Top Decay via FCNC}

\author{Zenr\=o Hioki\\
        Institute of Theoretical Physics,\ University of Tokushima,\\ Tokushima 770-8502, Japan\\
        E-mail: \email{hioki@tokushima-u.ac.jp}}

\author{\speaker{Kazumasa Ohkuma}\\
        Department of Information and Computer Engineering, Okayama University of Science,\\ Okayama 700-0005, Japan\\
        E-mail: \email{ohkuma@ice.ous.ac.jp}}

\author{Akira Uejima\\
        Department of Information and Computer Engineering, Okayama University of Science,\\ Okayama 700-0005, Japan\\
        E-mail: \email{uejima@ice.ous.ac.jp}}
\abstract{
We study possible non-standard $tuZ$ and $tcZ$ couplings, which induce flavor-changing 
neutral-current decays of the top quark, in the effective-Lagrangian framework. 
The corresponding interaction Lagrangian comes from several $SU(3) \times SU(2) \times U(1)$
invariant dimension-6 effective operators, and it includes four independent complex-number
coupling constants. Constraints on those non-standard coupling constants in each interaction
are derived by using the present experimental limits of the branching fractions for $t \to u Z$
and $t \to c Z$ processes. Expected improvements of the constraints at future facilities are
discussed as well. It is also pointed out that some correlations hold among those constrained
coupling constants.
}

\FullConference{XXIX International Symposium on Lepton Photon Interactions at High Energies - LeptonPhoton2019\\
		August 5-10, 2019\\
		Toronto, Canada}
\begin{document}
\section{Introduction}
Exploring rare decays of the top quark through precise measurements at the present and future 
facilities is quite challenging: Almost all the top quarks decay into the bottom quark and W-boson 
within the framework of the standard model. In particular, the event probability of $t \to qZ$ ($q=u/c$) 
processes caused by Flavor-Changing Neutral Current (FCNC) is so tiny there due to 
the Glashow-Iliopoulos-Maiani mechanism that it will be totally difficult to observe them even at future 
higher-energy\hspace{0.04cm}/\hspace{0.04cm}higher-luminosity experiments. Thus, once such $t\to qZ$ processes 
are discovered, it must 
be a strong indication of new physics beyond the standard model. Considering this situation, we recently 
have performed model-independent analyses of possible $tqZ$ couplings using the effective 
Lagrangian~\cite{Hioki:2018asl,Hioki:2019}. Here we would like to present main results of these works. 
\section{Analysis}
Assuming that there exists some new physics characterized by an energy scale ${\mit\Lambda}$ (e.g., the 
mass of a typical new particle) and all the non-standard particles are heavier than ${\mit\Lambda}$, 
the effective Lagrangian describing  $tqZ$ interactions around the electroweak scale is written as
\begin{alignat}{1}\label{eq:efflag_decay}
  &{\cal L}_{tqZ}  = -\frac{g}{2 \cos \theta_W} 
  \Bigl[\,\bar{\psi}_q(x)\gamma^\mu(f_1^L P_L + f_1^R P_R)\psi_t(x)Z_\mu(x) \Bigr.
 \nonumber\\
 &\phantom{===========}
  +\bar{\psi}_q(x)\frac{\sigma^{\mu\nu}}{M_Z}(f_2^L P_L + f_2^R P_R)
   \psi_t(x)\partial_\mu Z_\nu(x) \,\Bigr],
\end{alignat}
where $g$ is the $SU(2)$ coupling constant, $\theta_W$ is the weak mixing angle, $P_{L/R}\equiv(1\mp\gamma_5)/2$,
and $f_{1/2}^{L/R}$ are non-standard coupling constants including ${\mit\Lambda}$
(see Ref.~\cite{Hioki:2018asl,AguilarSaavedra:2008zc} for details).
These new coupling constants are in general complex numbers independent of each
other, and we do not impose any conditions on them from the viewpoint of model-independent study.
Thus, our analysis is carried out by varying the eight coupling parameters,
i.e., the real and imaginary parts of $f_{1/2}^{L/R}$.

The theoretical partial decay width ${\mit \Gamma}^{\rm th}_{tqZ}$ is therefore derived as an 
eight-variable function using the above Lagrangian. 
On the other hand, we have at present the following experimental information
(at 95 \% confidence level):
\begin{quote}\vspace*{-0.1cm}
$\bullet$ The total decay width of the top quark $\mit\Gamma^t$ [GeV]~\cite{Aaboud:2017uqq}
\vspace*{-0.2cm}
\begin{equation*}\label{eq:total_w}
4.8 \times 10^{-2} \leq {\mit\Gamma}^t \leq 3.5
\end{equation*}
%
$\bullet$ The upper limits of the branching fractions for $t\to qZ$ decays~\footnote{
     ``Future expectation'' means an assumption that ${\rm Br}(t \to qZ )$ will be
     reduced by half at future facilities.
}
\begin{center}\vspace*{-0.2cm}
\begin{tabular}{c|c} 
Current~\cite{ATLAS:2017beb} & Future expectation \\ \hline
 ${\rm Br}(t \to u Z ) < 1.7 \times 10^{-4}$&${\rm Br}(t \to u Z ) < 8.5 \times 10^{-5}$ \\
 ${\rm Br}(t \to c Z ) < 2.3 \times 10^{-4}$&${\rm Br}(t \to c Z ) < 1.2 \times 10^{-4}$
\end{tabular}
\end{center}
\end{quote}
\vspace*{-0.2cm}
Multiplying the minimum (maximum) value of ${\mit\Gamma}^t$ by ${\rm Br}(t \to u Z/c Z)$,
our input data ${\mit\Gamma_{tqZ}^{\rm exe}}$ [GeV]
are calculated as follows:\vspace*{-0.2cm}
\begin{center}
\begin{tabular}{c|c} 
Current                                                                        & Future expectation \\ \hline
 $0\leq{\mit\Gamma}_{tuZ}^{\rm exe} < 8.1 \times 10^{-6} ~~(5.9 \times 10^{-4})$&$0\leq{\mit\Gamma}_{tuZ}^{\rm exe} < 4.1 \times 10^{-6} ~~(3.0 \times 10^{-4})$ \\
 $0\leq{\mit\Gamma}_{tcZ}^{\rm exe} < 1.1 \times 10^{-5} ~~(8.0 \times 10^{-4})$&$0\leq{\mit\Gamma}_{tcZ}^{\rm exe} < 5.5 \times 10^{-6} ~~(4.0 \times 10^{-4})$
\end{tabular}
\end{center}
Then, in order to get allowed region of each $f_ {1/2}^{L/R}$,
a parameter space that satisfies 
${\mit \Gamma}^{\rm th}_{tqZ} < {\mit \Gamma}^{\rm exp}_{tqZ}$
is surveyed by varying the eight coupling parameters
at the same time.
\section{Results and Discussion}
Current constraints on the non-standard couplings in the $tuZ$ interactions are shown
in Table~\ref{tab:tuz_current} as one of the typical results:
\begin{table}[h]
\centering
\scriptsize
\caption{Current constraints on the non-standard couplings in the $tuZ$ interactions: those over (under) the dashed lines
in the rows denoted as Min. and Max. are the minimum and maximum of the allowed
ranges coming from ${\mit\Gamma}_{tuZ} = 8.1 \times 10^{-6} ~(5.9 \times 10^{-4})$.}
\label{tab:tuz_current}
\begin{tabular}{ccc|cc|cc|cc}
\multicolumn{1}{l}{\hspace*{-0.1cm}}                     & \multicolumn{2}{c|}{\hspace*{-0.1cm}$f_1^L$\hspace*{-0.1cm}}                                                                           & \multicolumn{2}{c|}{\hspace*{-0.1cm}$f_1^R$\hspace*{-0.1cm}}                                                                           &\multicolumn{2}{|c|}{\hspace*{-0.1cm}$f_2^L$\hspace*{-0.1cm}}                                                                           & \multicolumn{2}{c}{\hspace*{-0.1cm}$f_2^R$\hspace*{-0.1cm}}                                                                           \\ \cline{2-9} 
\multicolumn{1}{l}{\hspace*{-0.1cm}}                     & \hspace*{-0.1cm}Re($f_1^L$)\hspace*{-0.1cm}                                       & \hspace*{-0.1cm}Im($f_1^L)$ \hspace*{-0.1cm}                                       & \hspace*{-0.1cm}Re($f_1^R$)\hspace*{-0.1cm}                                       & \hspace*{-0.1cm}Im($f_1^R$)\hspace*{-0.1cm}                                       &\hspace*{-0.1cm}Re($f_2^L$)\hspace*{-0.1cm}                                       & \hspace*{-0.1cm}Im($f_2^L)$\hspace*{-0.1cm}                                        & \hspace*{-0.1cm}Re($f_2^R$)\hspace*{-0.1cm}                                       &\hspace*{-0.1cm} Im($f_2^R$) \hspace*{-0.1cm}                                      \\ \hline
\multicolumn{1}{c}{\multirow{2}{*}{\hspace*{-0.1cm}Min.\hspace*{-0.1cm}}}& \hspace*{-0.1cm}$-5.5\times 10^{-3}$\hspace*{-0.1cm}                              & \hspace*{-0.1cm}$-5.5\times 10^{-3}$\hspace*{-0.1cm}                               &\hspace*{-0.1cm} $-5.5\times 10^{-3}$\hspace*{-0.1cm}                              & \hspace*{-0.1cm}$-5.5\times 10^{-3}$\hspace*{-0.1cm}                              & \hspace*{-0.1cm}$-4.6\times 10^{-3}$\hspace*{-0.1cm}                              & \hspace*{-0.1cm}$-4.6\times 10^{-3}$\hspace*{-0.1cm}                               & \hspace*{-0.1cm}$-4.6\times 10^{-3}$\hspace*{-0.1cm}                              &\hspace*{-0.1cm} $-4.6\times 10^{-3}$ \hspace*{-0.1cm}                             \\ \cdashline{2-9} 
\multicolumn{1}{c}{\hspace*{-0.1cm}}                     & \multicolumn{1}{l}{\hspace*{-0.1cm}$-4.7\times 10^{-2}$\hspace*{-0.1cm}}          & \multicolumn{1}{l|}{\hspace*{-0.1cm}$-4.7\times 10^{-2}$\hspace*{-0.1cm}}          & \multicolumn{1}{l}{\hspace*{-0.1cm}$-4.7\times 10^{-2}$\hspace*{-0.1cm}}          & \multicolumn{1}{l|}{\hspace*{-0.1cm}$-4.7\times 10^{-2}$\hspace*{-0.1cm}}          &\multicolumn{1}{l}{\hspace*{-0.1cm}$-3.9\times 10^{-2}$\hspace*{-0.1cm}}          & \multicolumn{1}{l|}{\hspace*{-0.1cm}$-3.9\times 10^{-2}$\hspace*{-0.1cm}}          & \multicolumn{1}{l}{\hspace*{-0.1cm}$-3.9\times 10^{-2}$\hspace*{-0.1cm}}          & \multicolumn{1}{l}{\hspace*{-0.1cm}$-3.9\times 10^{-2}$\hspace*{-0.1cm}}         \\ \hline
\multicolumn{1}{c}{\multirow{2}{*}{\hspace*{-0.1cm}Max.\hspace*{-0.1cm}}}&\hspace*{-0.1cm} $\phantom{-}5.5\times 10^{-3}\hspace*{-0.1cm}$                    & \hspace*{-0.1cm}$\phantom{-}5.5\times 10^{-3}\hspace*{-0.1cm}$                     &\hspace*{-0.1cm} $\phantom{-}5.5\times 10^{-3}$\hspace*{-0.1cm}                    & \hspace*{-0.1cm}$\phantom{-}5.5\times 10^{-3}$\hspace*{-0.1cm}                    &\hspace*{-0.1cm}$\phantom{-}4.6\times 10^{-3}$\hspace*{-0.1cm}                    &\hspace*{-0.1cm} $\phantom{-}4.6\times 10^{-3}$  \hspace*{-0.1cm}                   & \hspace*{-0.1cm}$\phantom{-}4.6\times 10^{-3}$\hspace*{-0.1cm}                    & \hspace*{-0.1cm}$\phantom{-}4.6\times 10^{-3}$\hspace*{-0.1cm}                    \\ \cdashline{2-9} 
\multicolumn{1}{c}{\hspace*{-0.1cm}}                     & \multicolumn{1}{l}{\hspace*{-0.1cm}$\phantom{-}4.7\times 10^{-2}$\hspace*{-0.1cm}}& \multicolumn{1}{l|}{\hspace*{-0.1cm}$\phantom{-}4.7\times 10^{-2}$\hspace*{-0.1cm}}& \multicolumn{1}{l}{\hspace*{-0.1cm}$\phantom{-}4.7\times 10^{-2}$\hspace*{-0.1cm}}& \multicolumn{1}{l|}{\hspace*{-0.1cm}$\phantom{-}4.7\times 10^{-2}$\hspace*{-0.1cm}}&\multicolumn{1}{l}{\hspace*{-0.1cm}$\phantom{-}3.9\times 10^{-2}$\hspace*{-0.1cm}}& \multicolumn{1}{l|}{\hspace*{-0.1cm}$\phantom{-}3.9\times 10^{-2}$\hspace*{-0.1cm}}& \multicolumn{1}{l}{\hspace*{-0.1cm}$\phantom{-}3.9\times 10^{-2}$\hspace*{-0.1cm}}& \multicolumn{1}{l}{\hspace*{-0.1cm}$\phantom{-}3.9\times 10^{-2}$\hspace*{-0.1cm}}\\ \hline
\end{tabular}
\end{table}

\begin{figure}
\hspace*{-0.6cm}
\begin{minipage}{0.47\columnwidth}
\begin{itemize}
\item The allowed regions of the non-standard couplings in the $tqZ$ interactions are
given as $|f_{1/2}^{L/R}|<  O (10^{-3}) \sim O (10^{-2})$.
\vspace*{-0.6cm}
\item The non-standard couplings in the $tuZ$ interactions are more strongly restricted than those in the $tcZ$ interactions.
\vspace*{-0.1cm}
\item Both the real and imaginary parts of $f_{1}^{L/R}$ and $f_2^{L/R}$
in each of the $tuZ$ and $tcZ$ interactions have the same minimum and maximum limits respectively.
\vspace*{-0.6cm}
\item The allowed regions are expected to be narrowed by about 30\% 
if the branching fractions could be reduced by half at future facilities (e.g., High-Luminosity Large Hadron Collider).
\end{itemize}
\end{minipage}
\hspace*{0.45cm}
\begin{minipage}{0.52\columnwidth}
\includegraphics[bb=113 469 382 670,width=0.95\textwidth]{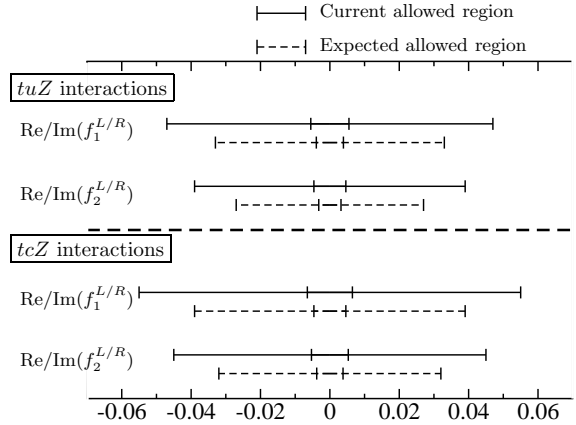}
\vspace*{0.05cm}\\
\caption{Current and expected constraints on the non-standard couplings in the $tuZ$ and $tcZ$ interactions.}
\label{fig1}
\end{minipage}
\end{figure}

\vspace*{0.2cm}
\noindent
All the other current and future expected constraints on 
the whole coupling parameters are listed in Ref.~\cite{Hioki:2018asl}.
Therefore, we here show only 
those results visually in Figure~\ref{fig1}:
The solid (dashed) lines mean the current (expected) allowed ranges for
the non-standard couplings in the $tuZ$ and $tcZ$ interactions.
The region between the two inside bars on each line is derived using 
${\mit\Gamma}_{tuZ} = 8.1 \times 10^{-6} (4.1 \times 10^{-6})$,
${\mit\Gamma}_{tcZ} = 1.1 \times 10^{-5} (5.5 \times 10^{-6})$  
and that between the two outside bars is derived using 
${\mit\Gamma}_{tuZ}=  5.9 \times 10^{-4} (3.0 \times 10^{-4})$,
${\mit\Gamma}_{tcZ} = 8.0 \times 10^{-4} (4.0 \times 10^{-4})$.
From Figure~\ref{fig1}, we can summarize the properties of 
the constrained coupling-parameters 
as follows:
\\ \vspace*{-0.4cm}

\begin{table}[t]
\centering
\scriptsize
\caption{Allowed minimum and maximum values of the $tuZ$ couplings for 
${\mit\Gamma}_{tuZ} = 8.1 \times 10^{-6}$ in the case that Re($f_1^L$) is fixed to
$5.5\times 10^{-3}$ which is the allowed maximum value.}
\label{tab:tuz_fixed}
\begin{tabular}{ccc|cc|cc|cc}
\multicolumn{1}{l}{\hspace*{-0.1cm}}     & \multicolumn{2}{c|}{\hspace*{-0.1cm}$f_1^L$\hspace*{-0.1cm}}                                          & \multicolumn{2}{c|}{\hspace*{-0.1cm}$f_1^R$\hspace*{-0.1cm}}                                     & \multicolumn{2}{c|}{\hspace*{-0.1cm}$f_2^L$\hspace*{-0.1cm}}                                          & \multicolumn{2}{c}{\hspace*{-0.1cm}$f_2^R$\hspace*{-0.1cm}}                                                      \\ \cline{2-9} 
\multicolumn{1}{l}{\hspace*{-0.1cm}}     &\hspace*{-0.1cm} Re($f_1^L$)\hspace*{-0.1cm}                          &\hspace*{-0.1cm} Im($f_1^L)\hspace*{-0.1cm}$                    &\hspace*{-0.1cm} Re($f_1^R$)\hspace*{-0.1cm}                    &\hspace*{-0.1cm} Im($f_1^R$)\hspace*{-0.1cm}                     &\hspace*{-0.1cm} Re($f_2^L$)\hspace*{-0.1cm}                          &\hspace*{-0.1cm} Im($f_2^L)$ \hspace*{-0.1cm}                   & \hspace*{-0.1cm}Re($f_2^R$) \hspace*{-0.1cm}                                    & \hspace*{-0.1cm}Im($f_2^R$)\hspace*{-0.1cm}                    \\ \hline
\multicolumn{1}{c}{\hspace*{-0.1cm}Min.\hspace*{-0.1cm}} & \vspace*{0.05cm}\multirow{2}{*}{\hspace*{-0.1cm}$5.5\times 10^{-3}$\hspace*{-0.1cm}} & \hspace*{-0.1cm}$-1.0\times 10^{-3}$\hspace*{-0.1cm}     & \hspace*{-0.1cm}$-1.0\times 10^{-3}$\hspace*{-0.1cm}           & \hspace*{-0.1cm}$-1.0\times 10^{-3}$\hspace*{-0.1cm}            &\hspace*{-0.1cm} $-8.0\times 10^{-4}$\hspace*{-0.1cm}                 &\hspace*{-0.1cm} $-8.0\times 10^{-4}$\hspace*{-0.1cm}           & \hspace*{-0.1cm}$-4.2\times 10^{-3}$\hspace*{-0.1cm}           & \hspace*{-0.1cm}$-8.0\times 10^{-4}$\hspace*{-0.1cm}           \\ \cline{1-1}\cline{3-9} 
\multicolumn{1}{c}{\hspace*{-0.1cm}Max.\hspace*{-0.1cm}} & \vspace*{0.05cm} {\hspace*{-0.1cm}\scriptsize (Fixed)\hspace*{-0.1cm}}               & \hspace*{-0.1cm}$\phantom{-}1.0\times 10^{-3}$\hspace*{-0.1cm} & \hspace*{-0.1cm}$\phantom{-}1.0\times 10^{-3}$\hspace*{-0.1cm} & \hspace*{-0.1cm}$\phantom{-}1.0\times 10^{-3}$\hspace*{-0.1cm}  & \hspace*{-0.1cm}$\phantom{-}8.0\times 10^{-4}$\hspace*{-0.1cm}       & \hspace*{-0.1cm}$\phantom{-}8.0\times 10^{-4}$\hspace*{-0.1cm} &\hspace*{-0.1cm}$-3.4\times 10^{-3}$\hspace*{-0.1cm}           & \hspace*{-0.1cm}$\phantom{-}8.0\times 10^{-4}$\hspace*{-0.1cm} \\ \hline
\end{tabular}
\end{table}
%

In addition to these studies, we also investigated if there is a certain relationship among the constrained 
couplings. For example,
when ${\rm Re}(f_1^L)$ is fixed to its maximum value and all the other constants are varied,
the allowed regions of the remaining couplings
are derived as Table~\ref{tab:tuz_fixed}.
From this table, we can see a relation between ${\rm Re}(f_1^L)$ and ${\rm Re} (f_2^R)$:
the sign of ${\rm Re} (f_2^R)$ is opposite to that of ${\rm Re}(f_1^L)$
and the size of ${\rm Re} (f_2^R)$ is the same order as ${\rm Re}(f_1^L)$.

As results of the similar analyses of the remaining couplings,
we found that the allowed region becomes the largest when there are relations as
${\rm Re/Im}(f_{1/2}^{L/R})=-C\,{\rm Re/Im}(f_{2/1}^{R/L})$ where
 $ 0.65 \lesssim C \lesssim 0.73$ ($0.93 \lesssim C \lesssim 1.1$) in the case that the maximum or minimum value
 of $f_1^{R/L} $ ($f_2^{R/L}$) is substituted in the right-hand side.
These correlations hold and work to maximize the allowed regions even 
if the non-standard couplings except for the negatively correlated two couplings
have no allowed regions. We confirmed that the results are common to the $tcZ$ couplings~(see Ref.\cite{Hioki:2019}
for detail discussions).
\section{Summary}
Non-standard couplings defined as $f_{1/2}^{L/R}$ in the effective Lagrangian describing FCNC $tqZ$ interactions
were studied as model-independently as possible. We firstly derived current and expected constraints 
on those coupling parameters based on available experimental information,
and then using the results we investigated if there exist any relationships among the constrained couplings.
It was found that the allowed region 
could be maximized when there is a specific correlation between ${\rm Re/Im}(f_{1/2}^{L/R})$ 
and ${\rm Re/Im}(f_{2/1}^{R/L})$.

Since our analyses are fully model independent, the results must be useful information to construct 
specific models with rather strong FCNC interaction.

\section*{Acknowledgement}
This work was partly supported by the Grant-in-Aid for Scientific Research (C) 
Grant Number 17K05426 from the Japan Society for the Promotion of Science.

\end{document}